**Scalable multilayer diffractive neural network with all-optical nonlinear activation**

Yiying Dong[1], Bohan Zhang[1], Ruiqi Liang[1,2], Wenhe Jia[1], Kunpeng Chen[1,2], Junye Zou[1,2], Futai Hu[1], Sheng Liu[3], Xiaokai Li[3], Yuanmu Yang[1]*

[1]State Key Laboratory of Precision Measurement Technology and Instruments, Department of Precision Instrument, Tsinghua University, Beijing 100084, China

[2]Weiyang College, Tsinghua University, Beijing 100084, China

[3]PICO

*ymyang@tsinghua.edu.cn

**All-optical diffractive neural networks (DNNs) offer a promising alternative to electronics-based neural network processing due to their low latency, high throughput, and inherent spatial parallelism. However, the lack of reconfigurability and nonlinearity limits existing all-optical DNNs to handling only simple tasks. In this study, we present a folded optical system that enables a multilayer reconfigurable DNN using a single spatial light modulator. This platform not only enables dynamic weight reconfiguration for diverse classification challenges but crucially integrates a mirror-coated silicon substrate exhibiting instantaneous $\chi^{(3)}$ nonlinearity. The incorporation of all-optical nonlinear activation yields substantial accuracy improvements across benchmark tasks, with performance gains becoming increasingly significant as both network depth and task complexity escalate. Our system represents a critical advancement toward realizing scalable all-optical neural networks with complex architectures, potentially achieving computational capabilities that rival their electronic counterparts while maintaining photonic advantages.**

**Introduction**

Neural network-based artificial intelligence has become essential in fields like machine vision and natural language processing[1-5]. With transistor miniaturization nearing physical limits, electronic hardware faces performance stagnation[6-8]. Optical computing using photons instead of electrons shows promise to significantly improve processing speed and computational throughput[9-15]. While optical neural networks (ONNs) were first conceptualized in the 1980s[16-18], their development stalled for decades due to inadequate optical neuron components and immature complex network research. Recent advances in ONNs employing Mach–Zehnder interferometers[19], micro-ring resonators[20,21], and diffractive optical elements[22] have revived interest. Particularly, free-space diffractive neural networks (DNNs) attract attention for their massive parallel computing capabilities[23,24].

A key limitation of free-space DNNs lies in their static optical components (Fig. 1a), which prevent dynamic weight adjustments that electronic neural networks easily achieve. Researchers have attempted using reconfigurable devices like digital micromirror devices[25,26], liquid crystal-based spatial light modulators (SLMs)[27,28], and active metasurfaces[29,30] (Fig. 1b) to solve this, but these solutions become impractical as networks grow deeper due to exponentially increasing component requirements.

Another major challenge is implementing nonlinear operations. While nonlinear activation is a fundamental component of neural networks, in most DNNs, the optical part performs only linear operations. Without interlayer nonlinearity, multiple hidden layers can be treated as a single linear layer[31,32]. The lack of nonlinear activation severely limits the performance of DNNs on more complex tasks.

One approach to mitigate the lack of optical nonlinearity is to develop optoelectronic hybrid neural networks, which use optics for linear computation and electronics for nonlinear activation[25,27,33]. However, the photoelectric conversion introduces additional latency and energy consumption. Recently, the data repetition method[34-36] has been proposed to realize nonlinear activation by exploiting scattering-induced nonlinear effects, eliminating the need for nonlinear optical materials. However, it requires repeating input data on the modulation layers, rendering it incapable of processing data that is not pre-registered using an electronic device.

Recently, there have been a number of efforts to enable all-optical nonlinear activation in DNNs (Fig. 1c). Atomic nonlinearities[37,38] have been explored, yet they require large setups and have a weak nonlinearity. Compact image intensifiers[39] can achieve optical-to-optical nonlinear activation but suffer from long response times (nanoseconds to milliseconds), undermining the DNNs' speed advantage[40]. Their fixed maximum light output also limits multilayer activations. Quantum dot photoluminescence[41,42] offers another approach, yet the inherent wavelength shift[43] also prohibits multilayer activations. These limitations hinder the development of fully optical, multilayer DNN systems scalable for handling various complex tasks.

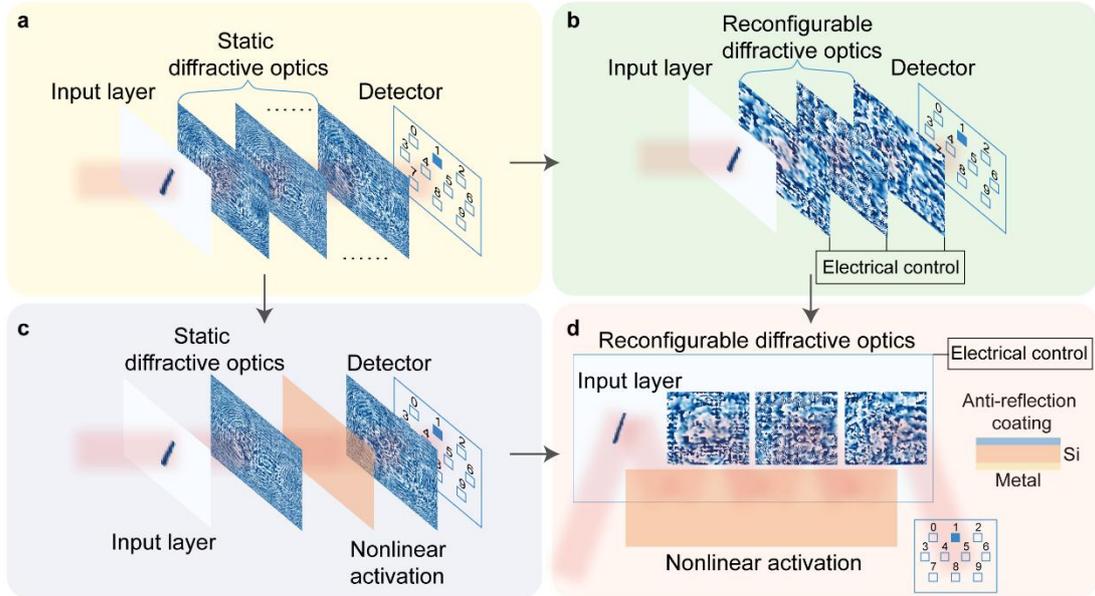

**Figure 1 | Comparison among different DNN configurations. a**, Schematic of a conventional DNN composed of static diffractive optical elements; **b**, Schematic of a reconfigurable DNN with electrically reconfigurable diffractive optical elements; **c**, Schematic of a DNN incorporating all-optical nonlinear activation; **d**, Schematic of the proposed scalable multilayer DNN using a single SLM for reconfigurable phase modulation and a mirror-coated silicon (Si) wafer for nonlinear activation.

In this study, we demonstrate a reconfigurable DNN system with multilayer all-optical nonlinear activation (Fig. 1d). Using a single SLM in a folded optical design, the system achieves $10^5$ programmable parameters distributed in 3 modulation layers. Nonlinear activation is enabled by the ultrafast $\chi^{(3)}$ optical nonlinearity of a mirror-coated 5-mm silicon (Si) wafer, which provides a full $2\pi$ phase shift while maintaining simplicity. The experimentally realized system can perform diverse tasks like classifying handwritten digits ("0" to "4" and "5" to "9"), and fashion products, with programmable weights in each diffractive layer. Our results further suggest that the incorporation of multilayer nonlinearity can facilitate a deeper DNN and offer a framework for building DNNs with more complex system architectures, with potential performance approaching that of modern electronics-based neural processors.

**Results**

**System configuration**. The proposed DNN system is schematically shown in Fig. 2a. It combines a phase-only SLM and a reflective nonlinear medium in a parallel configuration. This setup enables multiple phase modulations using a single SLM and achieves multilayer nonlinear activations through repeated reflections against the nonlinear medium. A Gaussian beam enters the SLM at a small angle of 3.55°, with input data loaded onto the first SLM block. The beam undergoes three stages of nonlinear activation and phase modulation before being captured by a camera for classification. To avoid interference from zero-order diffraction, the camera's detection zones are arranged in a ring-shaped pattern.

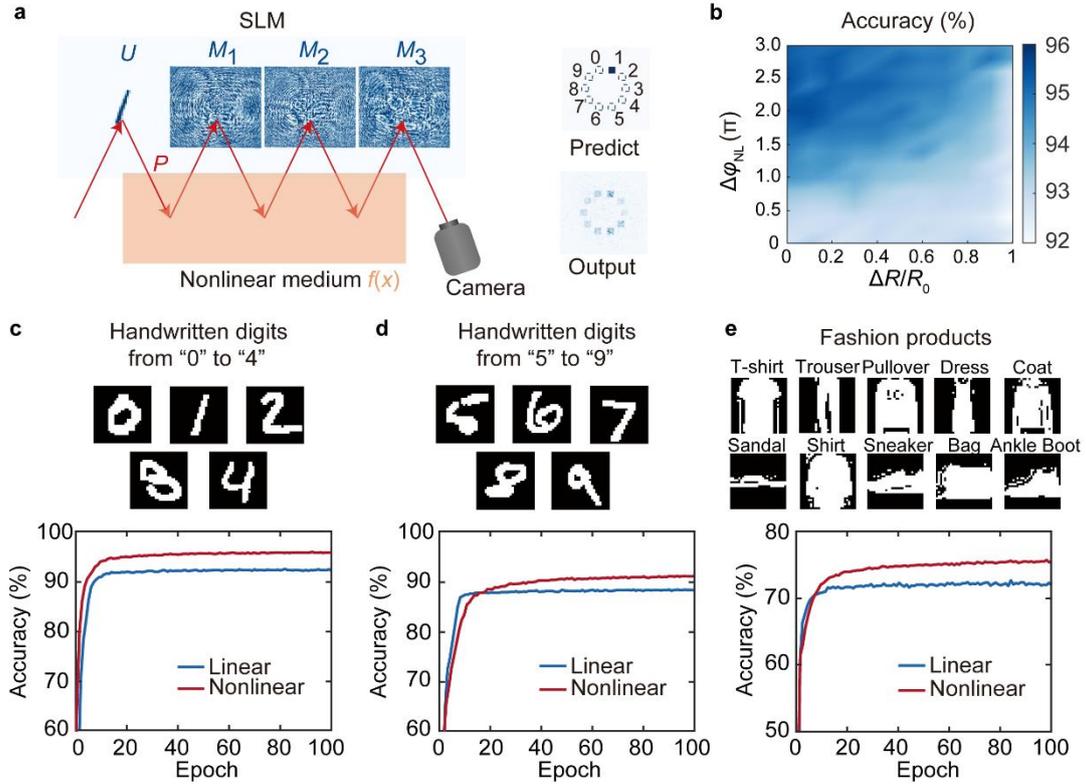

**Figure 2 | Design of the nonlinear reconfigurable DNN and impact of nonlinear activation. a**, Schematic of the nonlinear reconfigurable DNN. The designed detection regions on the camera sensor are arranged in a ring configuration to avoid unwanted zero-order diffraction. $U$, input field; $P$, propagation matrix; $f(x)$, nonlinear activation function; $M_N$, modulation matrix of the $N$th layer. **b**, Simulated classification accuracy for handwritten digits from "0" to "4" as a function of the nonlinear reflectance change and phase shift of the nonlinear medium. **c-e**, Comparison between simulated classification accuracy for DNNs with and without nonlinear activation for handwritten digits from "0" to "4" (**c**), handwritten digits from "5" to "9" (**d**), and 10-class fashion products (**e**).

The light propagation through the DNN can be described mathematically as,

$$T=PM_NPfP...(M_1Pf(P(U))),\qquad(1)$$

where $U$ represents the complex amplitude of the input field, $P$ is the free-space propagation matrix determined by the Rayleigh-Sommerfeld diffraction integral, $f(x)$ denotes the nonlinear activation function provided by the reflective nonlinear medium, $M_N$ corresponds to the modulation matrix of the $N$th layer, and $T$ is the output field with $|T|^2$ corresponding to the measured light intensity at the camera plane. $M_N$ is approximated to be independent of the incident angle since the phase response of the SLM is relatively insensitive to small incident angles, while $P$ is adjusted for the oblique incidence (Supplementary Section 1).

To establish the numerical model for $f(x)$, we first define the complex input field as $U = U_0 e^{j\varphi}$, where $U_0$ represents the amplitude and $\varphi$ denotes the phase. If we select a material with nearly instantaneous $\chi^{(3)}$ nonlinearity for low-latency activation, the nonlinear phase shift $\varphi_{NL}$ can be expressed as,

$$\varphi_{NL} = k\Delta nL = kn_2 UU^*L, \tag{2}$$

where $k$ is the wavenumber, $\Delta n$ is the change in refractive index, $n_2$ is the nonlinear refractive index, and $L$ is the nonlinear light-matter interaction distance. In addition, the two-photon absorption (TPA) process further induces a nonlinear reflectance change given by,

$$R_{NL} = R_0 e^{-\beta UU^*L}, \tag{3}$$

where $R_0$ is the linear reflectance and $\beta$ is the TPA coefficient. As a result, the nonlinear activation function combining reflectance change and phase shift is expressed as,

$$f(U) = \sqrt{R_{NL} \times UU^*} e^{j(\varphi + \varphi_{NL})}. \tag{4}$$

By incorporating the nonlinear activation function into Eq. (1), the forward model of the nonlinear DNN can be fully established, such that the weights of the DNN, which correspond to the phase values on the SLM, can be trained for given classification tasks via backpropagation (Methods and Supplementary Section 2).

To choose the proper nonlinear medium for activation, we first investigate the impact of nonlinear reflectance change and phase shift on the classification accuracy. Using a femtosecond laser operating at 1550 nm as the light source and with other system parameters designed to satisfy the fully connected condition[44] (Methods and Supplementary Section 3), we first simulated the DNN's classification accuracy of handwritten digits from "0" to "4" (a subset of the standard MNIST dataset) as a function of the nonlinear reflectance change and phase shift. The results, as shown in Fig. 2b, indicate that the classification accuracy is maximized when the maximum nonlinear phase shift approaches $2\pi$ with minimum nonlinear absorption. Further simulations confirmed that for various classification tasks including the classification of handwritten digits from "0" to "4", handwritten digits from "5" to "9", and fashion products from the Fashion-MNIST dataset, the classification accuracy can all be largely enhanced after incorporating $2\pi$ nonlinear phase shift and zero nonlinear reflectance change, as shown in Fig. 2c-e.

Using the results presented in Fig. 2 as a guideline, we surveyed existing nonlinear optical materials with large $n_2$ and small $\beta$ (Supplementary Section 4). While semiconductor saturable absorber mirrors are widely used $\chi^{(3)}$ media, they suffer from strong nonlinear absorption and weak nonlinear phase shifts[45,46]. Emerging materials like epsilon-near-zero materials[47-49] and 2D materials[50] exhibit high effective $n_2$, but their short light-matter interaction distances (nanometer/micrometer scale) limit total nonlinear phase accumulation. Si was ultimately selected as the nonlinear medium. As a mature semiconductor, Si offers a large nonlinear refractive index $n_2$ of $4.5 \times 10^{-18}$ m$^2$/W at the wavelength of 1550 nm[51,52]. With negligible linear absorption and moderate nonlinear absorption, light can interact with Si over an extended distance in the millimeter range, thus allowing it to achieve a large total nonlinear phase shift, making it an ideal candidate for nonlinear activation in DNNs.

**Experimental results**. The fabricated nonlinear medium involved a double-side-polished 5-mm-thick Si wafer with a metallic mirror coating on the backside (to enhance reflection) and an anti-reflection coating on the front side. This configuration achieved a measured reflectance of 90.44% at 1550 nm. To calibrate the nonlinear response, we set up a system as depicted in Fig. 3a. For the nonlinear phase shift measurement, a 1550 nm femtosecond laser beam with a Gaussian cross-section was split by a 50:50 beam splitter, with one beam directed at the mirror-coated Si wafer and the other at a reference metallic mirror. Interference fringes from the combined beams were analyzed to quantify phase shifts versus input power (Supplementary Section 5). Meanwhile, the nonlinear reflectance change was measured using the I-scan technique[53]. Although the measured maximum nonlinear phase shift of the mirror-coated Si wafer (shown in the right panel of Fig. 3a) does not reach the full 2π range with the Gaussian incident beam, in actual classification experiments, light modulated by the SLM without passing through the beam splitter can have a higher local intensity, resulting in a larger nonlinear phase shift. The measured nonlinear response of the mirror-coated Si wafer was subsequently fed into the forward model of the nonlinear DNN to re-train the phase values on the SLM for different classification tasks using backpropagation (Supplementary Section 6).

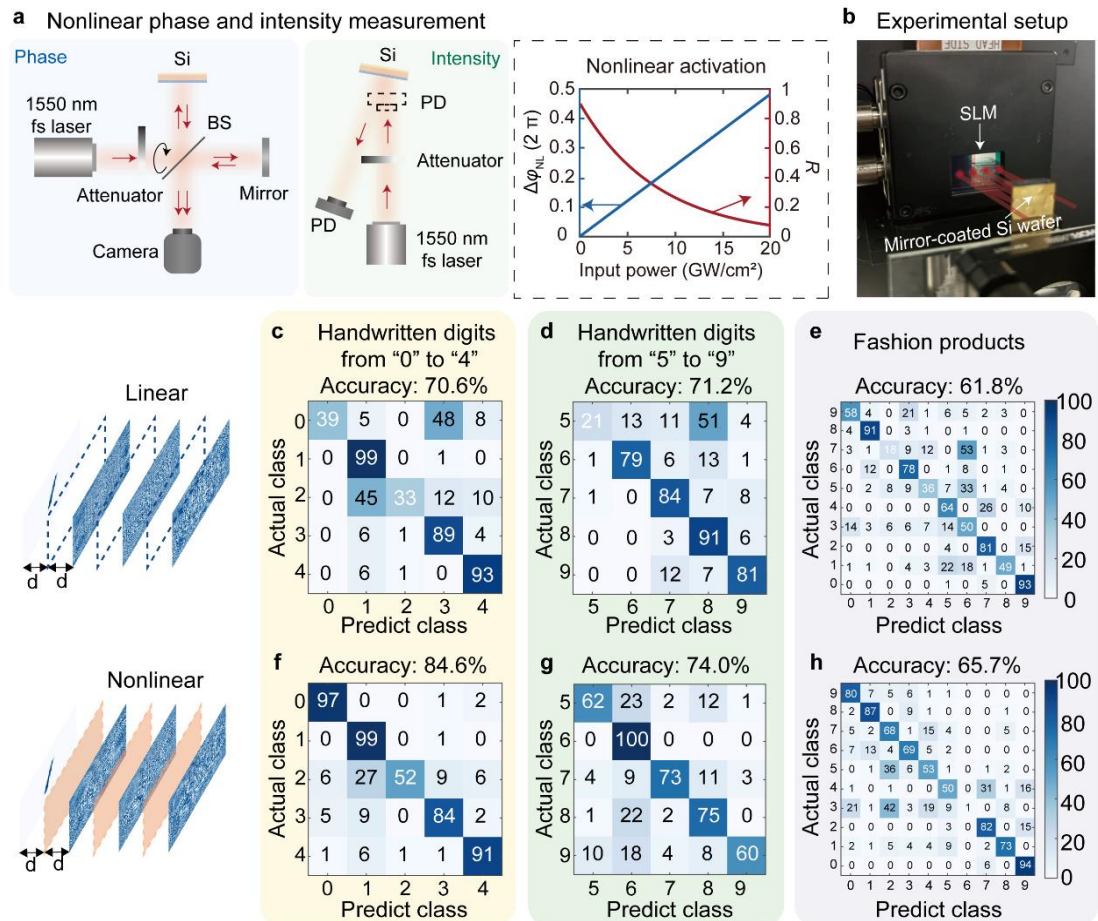

**Figure 3 | Experimental characterization of the nonlinear reconfigurable DNN. a**, Schematic of the experimental setup for the nonlinear phase shift (left panel) and reflectance change (middle panel) measurement. The right panel shows the measured results. BS, beam splitter; Si, mirror-

coated Si wafer; PD, photodetector. **b**, Photograph of the nonlinear reconfigurable DNN. **c-e**, Confusion matrices of the linear DNN for classifying handwritten digits from "0" to "4" (**c**), handwritten digits from "5" to "9" (**d**), and fashion products (**e**). **f-h**, Confusion matrices of the nonlinear DNN for classifying handwritten digits from "0" to "4" (**f**), handwritten digits from "5" to "9" (**g**), and fashion products (**h**).

Figure 3b shows a photograph of the experimental setup for the DNN demonstration (details in Methods). To demonstrate the DNN's reconfigurability, three classification tasks were tested: handwritten digits ("0" to "4" and "5" to "9") and fashion products. Using 100 test samples per category under low-power laser illumination (nonlinear effects disabled), the confusion matrices in Fig. 3c-e show classification accuracies of 70.6%, 71.2%, and 61.8%, respectively.

To assess the impact of nonlinear activation, the laser peak intensity was increased to 6.33 GW/cm². As shown in Fig. 3f-h, classification accuracy improved significantly with nonlinear activation. While experimental accuracy for both linear and nonlinear cases was lower than simulations due to forward model errors, the experimental improvement in accuracy with nonlinear activation surpassed simulations. This suggests that the nonlinear system is more robust against system errors than the linear system (Supplementary Section 7).

**Nonlinearity-assisted scalable DNN.** Most existing all-optical DNNs struggle with complex tasks due to the absence of nonlinear activation. In our experiment, we also only implemented three modulation layers and performed simple classification tasks due to the limited pixel number of the SLM. However, the folded geometry with multilayer nonlinear activation enables straightforward scaling for more complex tasks, given that there is no fundamental limit to increasing the pixel number of a liquid crystal-based SLM.

To assess scalability, we compared DNN performance as a function of the modulation layer with and without nonlinear activation across progressively complex classification tasks: 5-class handwritten digits from "0" to "4", 10-class Fashion-MNIST, and 25-class QuickDraw. The results, as shown in Fig. 4a, suggest that the classification performance of a nonlinear DNN can be greatly boosted compared to a linear DNN, particularly in deeper networks for complex tasks. For complex tasks such as the classification of the 25-class QuickDraw dataset, the 9-layer nonlinear DNN achieves a notable accuracy improvement of 26.33%.

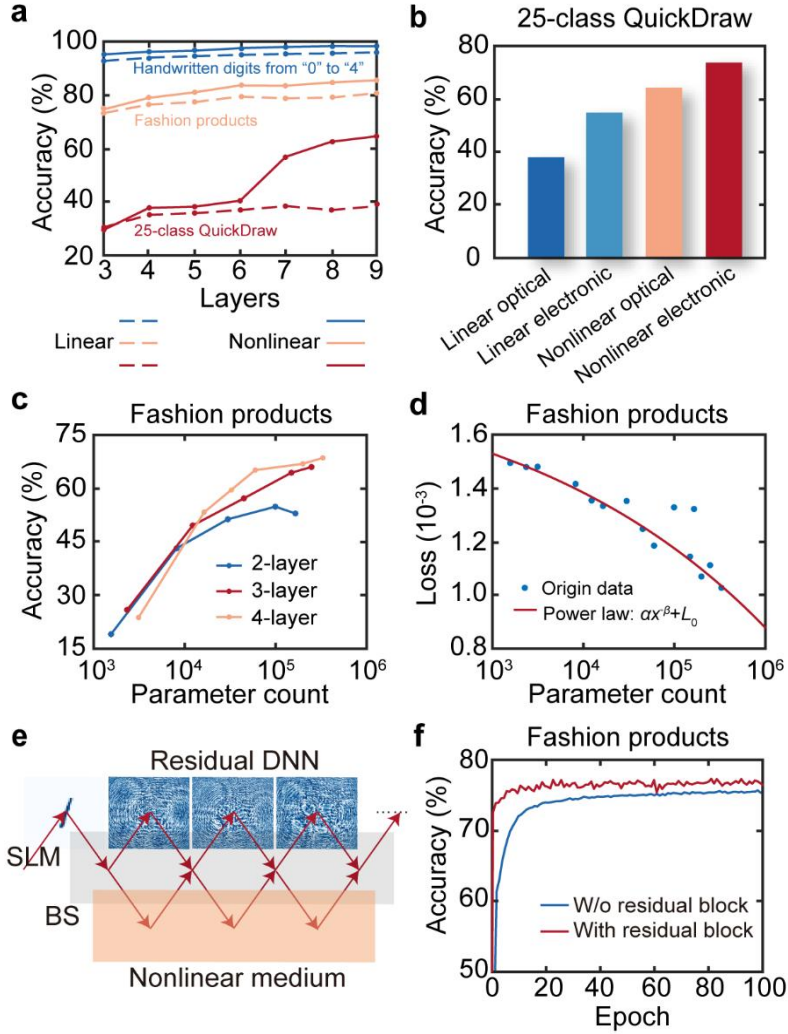

**Figure 4 | Evaluation of the scalability of a nonlinear DNN. a**, Comparison of classification accuracy with and without nonlinear activation as a function of the layer number for different classification tasks of handwritten digits from "0" to "4", fashion products, and 25-class QuickDraw dataset. **b**, Comparison of classification accuracy for the 25-class QuickDraw dataset using a linear DNN, a linear electronic multilayer perceptron (MLP), a nonlinear DNN, and a nonlinear electronic MLP with a similar number of parameters. **c**, Comparison of fashion products classification accuracy as a function of parameter count with 2, 3 and 4 layers, respectively. In the scalability test, the original 28×28 image size of the fashion products dataset was used to ensure good performance with smaller modulation units, whereas other simulations used upscaled 168×168 input images for improved accuracy. **d**, Comparison of testing loss as a function of parameter count. The original data is fitted to a power function $Loss = \alpha x^{\beta} + L_0$, where $\alpha, \beta, L_0$ are fitting parameters and $x$ represents parameter count. **e**, Schematic of the scalable nonlinear DNN with residual connection realized by a single beam splitter. **f**, Comparison of fashion products classification accuracy using 3-layer nonlinear DNNs with and without residual blocks, respectively.

By integrating multilayer nonlinear activation, the nonlinear DNN's performance can potentially approach that of electronic neural processors. For the 25-class QuickDraw dataset classification, we trained a 9-layer digital multilayer perceptron (MLP) with and without nonlinear activation, containing 736,818 parameters (nearly matching the DNN's 736,164 parameters). As shown in Fig. 4b, the nonlinear DNN achieved 64.53% accuracy, surpassing both linear DNN (38.20%) and linear MLP (55.23%). However, compared to the digital MLP with ReLU activation (74.10% accuracy), the DNN's lower performance (64.53%) likely stems from its limitation to non-negative weight values. The performance of the nonlinear DNN can be further enhanced by scaling up parameter counts, as demonstrated by the power-law relationship between testing loss and parameter quantity (Fig. 4c-d) that follows the same trend of digital deep MLP networks[54,55]. Additionally, architectural modifications like incorporating residual connections[56] through a simple beam splitter (Fig. 4e) can further improve the DNN's performance: a 3-layer DNN with residual blocks achieves 77.20% accuracy on Fashion-MNIST, outperforming its non-residual counterpart (75.46%) as shown in Fig. 4f (Supplementary Section 8).

**Discussion**

In summary, we introduced a reconfigurable DNN system that integrates all-optical nonlinear activation using a single SLM for multilayer modulation and a mirror-coated Si wafer as the nonlinear component. This minimalist design achieved high performance in classifying handwritten digits ("0" to "4" and "5" to "9") and 10-class fashion products, with the nonlinear DNN largely surpassing its linear counterpart. Accuracy could be further enhanced through online training methods like digital twins[57], physics-aware algorithms[58], or camera-in-the-loop optimization[59]. By scaling SLM pixel density and architectural complexity, the system's capacity can expand to tackle more sophisticated tasks.

The main drawback of the proposed nonlinear DNN is that it still requires a high-power femtosecond laser for nonlinear activation. This limitation could be mitigated by developing advanced nonlinear optical materials, with our findings highlighting that the ideal materials for DNN nonlinear activation should combine a high nonlinear refractive index with minimal linear/nonlinear absorption. Additionally, extending the Si $\chi^{(3)}$ nonlinearity approach to silicon-on-insulator (SOI) chip-based DNNs[60] could enhance nonlinear effects under identical power levels due to increased energy density. Beyond conventional image classification, the system shows unique potential for femtosecond laser-based applications, such as analyzing spatiotemporal laser beam profiles.

**Methods**

**Datasets and training details.** We chose MNIST, Fashion-MNIST, and QuickDraw datasets to benchmark the performance of the DNN. The MNIST dataset was split into two subsets: one containing handwritten digits from "0" to "4", and the other containing handwritten digits from "5" to "9". The first subset (digits from "0" to "4") consists of 15298 training images, 15298 validation images, and 4861 test images, while the other one (digits from "5" to "9") contains 14702 training images, 14702 validation images, and 5139 test images. The Fashion-MNIST dataset consists of 30,000 training images, 30000 validation images and 10,000 test images across 10 different categories of fashion products. The QuickDraw dataset is a collection of over 50 million drawings across 345 categories, contributed by players of the game "Quick, Draw!". We selected 25 classes, using 2000 images per class for training, 2000 for validation, and 1000 for testing (Supplementary Section 9).

     The DNN was modeled numerically using TensorFlow (v2.10.0, Google Inc.) on a server equipped with Nvidia A100 GPU. We trained the model for each task using the mean square error loss as a loss function to quantitatively describe the differences between the designed distribution

and the actual output on the sensor plane. The phase profile in each layer of the network was updated by a stochastic gradient descent algorithm with a training batch size of 16 and a learning rate of 0.002, with the desired mapping function between the input and output planes achieved after 100 epochs.

**Experimental setup.** The experimental setup for characterizing the DNN is shown in Fig. 3b. The light source used was an optical parametric amplifier (Light Conversion Orpheus ONE-HP) operating at a wavelength of 1550 nm, pumped by a Yb: KGW laser oscillator (Light Conversion Pharos). The light was directed onto a phase-only SLM (Hamamatsu X15213-15L), which has 1272×1024 modulation elements, with a 12.5-μm pixel size and an 8-bit accuracy, for input information loading and phase modulation. The laser power was kept below the SLM's damage threshold as instructed by engineers from Hamamatsu Inc.. Finally, the image is captured by an InGaAs camera (Hamamatsu C14041-10U) after attenuation.

The incident laser power was adjusted by variable neutral density filters (Thorlabs NENIR10A-C and NENIR40A-C). For the linear DNN, the peak intensity of the laser beam is reduced to 133 $kW/cm^2$, with a beam diameter of 3.6 mm. For the nonlinear DNN, the peak intensity of the laser beam is increased to 6.33 $GW/cm^2$.

**Data availability**

All relevant data are available in the main text, in the Supporting Information, or from the authors.

**Code availability**

The source code is available at https://github.com/THUMetaOptics/NDNN.

**Acknowledgment**

This work was supported by the National Natural Science Foundation of China (62135008) and by ByteDance.

**Author contributions**

Y.Y., F.H. and Y.D. conceived this work. Y.D. developed the digital neural network and trained the phase masks; R.L., B.Z., K.C. and J.Z. optimized the angular spectrum method; W.J. assisted with the selection and characterization of the nonlinear medium. Y.D. conducted the experiment; Y.D. and Y.Y. wrote the manuscript with input from other authors. Y.Y. supervised the project.

**Competing interests**

The authors declare no competing interests.

**Supplementary Information:**

**Scalable multilayer diffractive neural network with all-optical nonlinear activation**

Yiying Dong[1], Bohan Zhang[1], Ruiqi Liang[1,2], Wenhe Jia[1], Kunpeng Chen[1,2], Junye Zou[1,2], Futai Hu[1], Sheng Liu[3], Xiaokai Li[3], Yuanmu Yang[1]*

[1]State Key Laboratory of Precision Measurement Technology and Instruments, Department of Precision Instrument, Tsinghua University, Beijing 100084, China

[2]Weiyang College, Tsinghua University, Beijing 100084, China

[3]PICO

*ymyang@tsinghua.edu.cn

## 1. Derivation of the propagation matrix for the oblique incidence

To cascade multiple modulation layers on the spatial light modulator (SLM), the light is directed to be incident onto the SLM at a small angle. As a result, the propagation matrix $P$ in Eq. (1) of the main text must be modified to account for this oblique incidence. For efficient computation, the angular spectrum method is used to calculate the diffraction process. The propagation matrix that transforms the incident light can be expressed as,

$$P(u(x_0, y_0)) = u(x, y) = u(x_0, y_0) \otimes \frac{\exp(jkd)}{jkd} \exp(j\frac{k}{2d}(x^2 + y^2)), \tag{S1}$$

where $u(x, y)$ is the output field and $u(x_0, y_0)$ is the input field. The propagation distance $d$ is defined as $z$-$z_0$ and $\otimes$ symbolizes the convolution operator with $k$ being the wavenumber. To streamline the convolution calculation, the operation is transformed from the spatial domain into the spatial frequency domain, where it simplifies to multiplication as,

$$A(f_x, f_y) = F(u(x_0, y_0)) \times H(f_x, f_y), \tag{S2}$$

$$u(x, y) = F^{-1}(A(f_x, f_y)), \tag{S3}$$

where $A(f_x, f_y)$ represents the spatial frequency spectrum of the output field, with $F$ and $F^{-1}$ denoting the Fourier and inverse Fourier transform, respectively. The transfer function in the spatial frequency domain $H(f_x, f_y)$ is given by,

$$H(f_x, f_y) = F(\frac{\exp(jkd)}{jkd} \exp(j\frac{k}{2d}(x^2 + y^2))) = \exp(jkd\sqrt{1 - \lambda^2 f_x^2 - \lambda^2 f_y^2}), \tag{S4}$$

where $\lambda$ is the operating wavelength. Thus, the output field is computed as,

$$u(x, y) = F^{-1}(F(u(x_0, y_0)) \times H(f_x, f_y)). \tag{S5}$$

For oblique incidence, two key adjustments are required. First of all, the center frequency of the spatial frequency spectrum shifts from $f_{x\_ori} = f_{y\_ori} = 0$ to the frequency $f_{x\_cor} = 0$, $f_{y\_cor} = \sin\theta_{in}/\lambda$, where $\theta_{in}$ is the incident angle. Secondly, the output field center moves from $(x_{ori}, y_{ori}) = (0,0)$ to $(x_{cor}, y_{cor}) = (0, d\tan\theta_{in})$. The corrected transfer function is written as,

$$H_{cor}(f_x, f_y) = \exp(jkd\sqrt{1 - \lambda^2 f_x^2 - \lambda^2 (f_y - f_{y\_cor})^2}) + j2\pi f_y y_{cor}. \tag{S6}$$

The transformation of the propagation matrix for oblique incidence is expressed as,

$$P(u(x_0, y_0)) = u(x', y') = F^{-1}(F(u(x_0, y_0)) \times H_{cor}(f_x, f_y)), \tag{S7}$$

where $x' = x$-$x_{cor}$ and $y' = y$-$y_{cor}$.

## 2. Training process of the diffractive neural network

The training process of the diffractive neural network (DNN) is schematically illustrated in Fig. S1. A Gaussian beam is incident onto the first block of the SLM, encoding the input information. After three stages of nonlinear activation and phase modulation, the output is captured by an infrared camera for classification. To minimize the influence of zero-order diffraction, the detector regions are arranged in a ring configuration. The mean-squared error (MSE) loss quantifies the difference between the designed and actual output distributions, defined as,

$$\text{MSE} = \frac{1}{M}\sum_{m=1}^{M}(x_m\text{-}y_m)^2, \tag{S8}$$

where $M$ is the total number of pixels on the output plane, $x_m$ is the predicted value at point $m$, and $y_m$ is the actual value. The parameters are iteratively updated via backpropagation[1], generating optimized phase masks on the SLM for different classification tasks. Finally, the parameters are iteratively updated using the backpropagation algorithm, resulting in different phase masks on the SLM for different classification tasks.

In the current configuration, a single SLM is used for input encoding and phase modulation to simplify the hardware configuration. However, this causes the input field to undergo nonlinear activation before phase modulation, differing from conventional neural networks. This issue can be resolved by employing separate SLMs for input encoding and phase modulation.

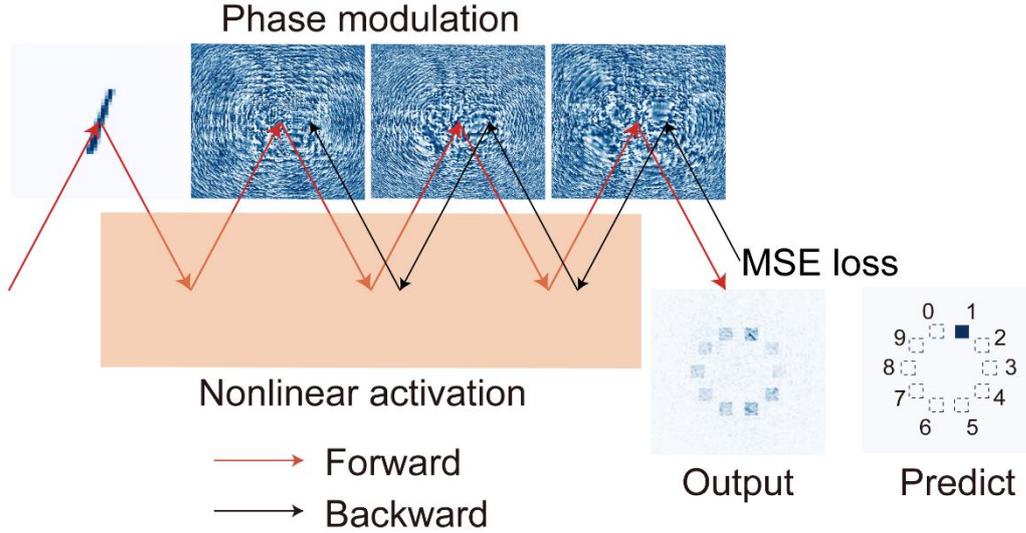

**Figure S1** | Training process of the DNN. The red and black arrows represent the diffractive propagation of the forward model and the backpropagation-based training process, respectively.

## 3. Parameter design for the folded nonlinear DNN system

To optimize classification performance and fully utilize the SLM, we designed the experimental system based on the fully connected condition[2]. According to the diffractive optics theory, the maximum diffraction angle $\varphi_{max}$ of a diffractive optical element (DOE) is given by,

$$\varphi_{max} = \sin^{-1}\left(\frac{\lambda}{2a}\right), \tag{S9}$$

where $\lambda$ is the operating wavelength and $a$ is the pixel size of the DOE. To satisfy the fully connected condition and maximize the SLM utilization, the incident angle $\theta_{in}$ is set to be equal to $\varphi_{max}$, as shown in Fig. S2b. The distance $D$ between the front surface of the mirror-coated silicon (Si) wafer and the SLM must satisfy,

$$2\times(D\tan\theta_{\text{in}} + L_{\text{n}}\tan\theta_2) = \sqrt{N_{\text{SLM}}}\times a, \tag{S10}$$

where $L_{\text{n}}$ is the thickness of the mirror-coated Si wafer, $N_{\text{SLM}}$ is the number of pixels per modulation layer, and $\theta_2$ is the refraction angle inside the mirror-coated Si wafer, determined by,

$$n_{\text{air}}\sin\theta_{\text{in}} = n_{\text{Si}}\sin\theta_2, \tag{S11}$$

where $n_{\text{air}}$ and $n_{\text{Si}}$ are the refractive indices of air and Si, respectively.

With an operating wavelength of 1550 nm and an SLM pixel size of 12.5 μm, the incident angle is set to be 3.55°, as determined by Eq. (S9). Each layer consists of 298×298 pixels, with 286×286 used for modulation, leaving a 12-pixel gap between layers to prevent light overlap. The Si wafer thickness is 5 mm. According to Eq. (S10), the distance $D$ is calculated as 2.86 cm. To facilitate camera mounting, the distance between the SLM's output surface and the camera ($D_{\text{cam}}$) is set to 15 cm.

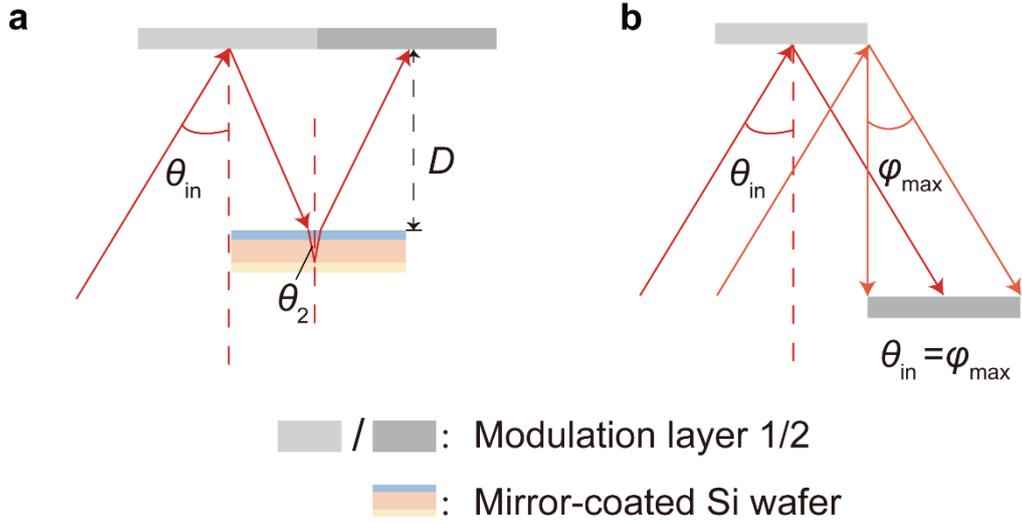

**Figure S2 |** Illustration of the fully connected condition for the system parameter design. **a,** Schematic of a portion of the folded nonlinear DNN system, illustrating a single reflection scenario. **b,** Unfolded transmission counterpart of the folded system shown in panel **a**, providing an equivalent perspective. $\theta_{\text{in}}$, incident angle; $\theta_2$, angle of refraction in the Si wafer; $D$, the distance between the SLM and the front surface of the mirror-coated Si wafer; $\varphi_{\text{max}}$, maximum diffraction angle.

## 4. Comparison of nonlinear properties of different $\chi^{(3)}$ materials

To clarify our choice of Si as the activation material, we present the nonlinear parameters of representative $\chi^{(3)}$ materials in Table S1. Si provides a significant nonlinear phase shift and moderate reflectance change under the incident power of 1 GW/cm², which aligns with the requirements depicted in Fig. 2b of the main text. At the same time, the linear absorption of Si is relatively low. These properties make Si well-suited for nonlinear activation. Furthermore, by shifting the operating wavelength from 1550 nm to 2200 nm, a larger nonlinear phase shift can be achieved with minimum two-photon absorption[3], potentially further improving the system performance.

**Table S1  Comparison of nonlinear properties of different $\chi^{(3)}$ materials[a]**

| Materials | $\lambda$ (nm) | $n_{2eff}$ (cm²/GW) | $\beta_{eff}$ (cm/GW) | $L$ (cm) | $\Delta\varphi$ (2π) @ 1 GW/cm² | $\Delta R/R_0$ (%) @ 1 GW/cm² | $A_0$ (%) |
|---|---|---|---|---|---|---|---|
| SESAM[b] | 1550 | / | / | / | / | $\Delta R$ =34.00%[c] | 21.00 |
| Few-layer graphene[4,5d] | 532 | 60[e] | / | ~1×10⁻⁷ | 0.11 | / | / |
| Single-layer graphene[6,5] | 733 | 1.4 | 6×10³ | 3.3×10⁻⁸ | 6.30×10⁻⁴ | +0.02 | 2.30 |
| ITO[7] | 1250 | 0.11 | 7.13×10³ | 3.1×10⁻⁵ | 2.73×10⁻² | +2.25 | / |
| Carbon nanotube[8] | 1460 | 1.1×10⁻² | 5.4×10² | 4.2×10⁻⁵ | 3.16×10⁻³ | +2.24 | / |
| ZnSe[9] | 532 | 6.9×10⁻⁵ | 5.8 | 0.27 | 0.35 | -79.11 | 21.3[10] |
| Si[3e] | 1550 | ~4.5×10⁻⁵ | ~1.50 | 1 | 0.29 | -77.69 | 0[11] |
| Si[3] | 2200 | ~9×10⁻⁵ | ~0.4 | 1 | 0.41 | -32.97 | 0[11] |

[a]$\lambda$, operating wavelength; $n_{2eff}$, effective nonlinear refractive index; $\beta_{eff}$, effective two-photon absorption coefficient; $L$, thickness of the $\chi^{(3)}$ material; $\Delta\varphi$, nonlinear phase shift induced by an incident intensity of 1 GW/cm², neglecting absorption; $\Delta R/R_0$, nonlinear reflectance/transmittance change, defined as: $\Delta R/R_0 =1\text{-exp}(-\beta_{eff}I_{2\pi}L)$, "+" denotes saturable absorption, "-" denotes reverse saturable absorption. $A_0$, linear absorbance.

[b]Data corresponding to SAM-1550-55-2ps-x derived from https://www.batop.de/products/saturable-absorber/saturable-absorber-mirror/saturable-absorber mirror-1550nm.html.

[c]$\Delta R$ is the maximum modulation depth of SESAM at 1550 nm, with a saturation fluence of 40 μJ/cm².

[d]The nonlinear refractive index of graphene varies with the input power.

[e]**The Si thickness ($L$) matches the experimentally used Si wafer. Other linear and nonlinear parameters were taken from Refs. [3, 11].**

## 5.  Fabrication and characterization of the mirror-coated Si wafer

Experimentally, we deposited a metallic coating on the backside of a double-side-polished 5-mm-thick Si wafer to enhance the back side reflection and an anti-reflection coating on the front side of the Si wafer to minimize the front side reflection, as shown in Fig. S3a. The metallic layer consists of 300 nm aluminum (Al), 50 nm titanium (Ti), and 50 nm gold (Au), while the anti-reflection film is composed of silicon dioxide ($SiO_2$) and titanium dioxide ($TiO_2$). The linear reflectance of the mirror-coated Si wafer reaches 90.44% at 1550 nm, with the loss primarily attributed to the absorption and scattering induced by the metallic coating.

To calibrate the nonlinear phase shift and reflectance change of the mirror-coated Si wafer for the DNN demonstration, we set up the experimental system as shown in Fig. 3a of the main text. For the nonlinear phase shift measurement, we recorded the variation in the interference fringes as the laser input power was varied, as shown in Fig. S3b. As the laser peak intensity increases from 0.01 GW/cm$^2$ to 3.27 GW/cm$^2$, a significant shift in the peak of the stripes occurs, allowing the nonlinear phase shift to be calculated.

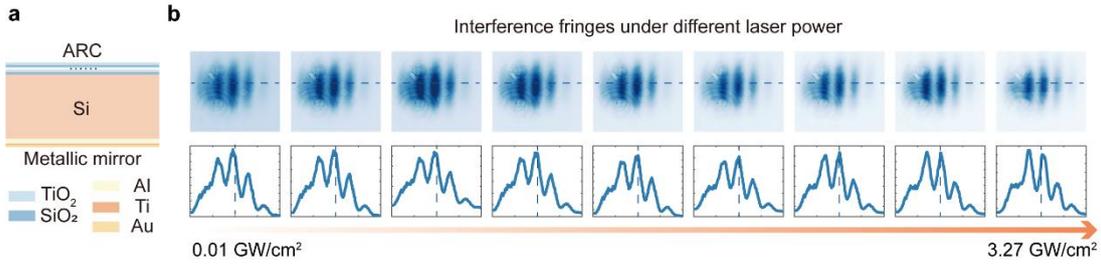

**Figure S3 | a**, Schematic of the mirror-coated Si wafer: The metallic mirror consists of Al, Ti, and Au layers, while the anti-reflection coating is composed of $SiO_2$ and $TiO_2$. ARC, anti-reflection coating. **b**, Interference fringes change as a function of the input power. The upper panel shows the original interference fringes, while the bottom panel displays the intensity distribution along the dashed line in the original fringes.

## 6. Simulation results refined by the measured nonlinear response of the mirror-coated Si wafer

To improve the consistency between the experimental results and simulations, we correct the simulation by incorporating the measured nonlinear parameters of the mirror-coated Si wafer. As shown in Fig. S4, the nonlinear activation effect is less pronounced than in the ideal case (Fig. 2c-e of the main text). This discrepancy occurs because, although the nonlinear phase shift approaches $2\pi$, the substantial change in nonlinear reflectance leads to a drop in accuracy, as indicated by the point (0.9,1) in Fig. 2b of the main text, where a relatively small accuracy improvement is observed.

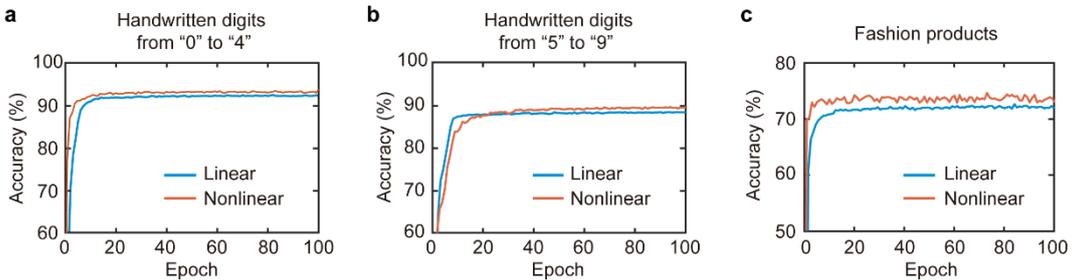

**Figure S4 |** Comparison of classification accuracy with and without nonlinear activation using the measured nonlinear parameters of the mirror-coated Si wafer for handwritten digits from "0" to "4" (**a**), handwritten digits from "5" to "9" (**b**), and 10-class fashion products (**c**).

## 7. Robustness against system error for the nonlinear DNN

Experimentally, we observed a more pronounced improvement in classification accuracy for the nonlinear DNN compared to the linear DNN. To explore this phenomenon, we simulated the classification accuracy of both linear and nonlinear DNNs for handwritten digits from "0" to "4", while varying potential system errors. These errors included the distance between the SLM and the mirror-coated Si wafer and the incident angle, with the modulation phase kept constant as in the error-free case. As shown in Fig. S5, the nonlinear DNN consistently maintains higher accuracy than the linear system across varying errors, indicating that the nonlinear DNN is more tolerant to system errors. The robustness may be attributed to the fact that the nonlinear activation function allows higher-level abstractions of the input data and better control of the gradient magnitude during backpropagation, ultimately leading to a more resilient model[12].

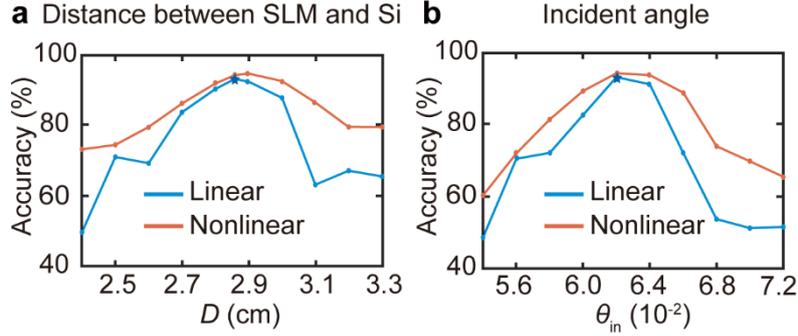

**Figure S5 |** Comparison of the classification accuracy of the linear and nonlinear DNN as a function of the distance between the SLM and the front side of the mirror-coated Si wafer $D$ (**a**) and the incident angle $\theta_{in}$ (**b**). The stars indicate the error-free cases.

## 8. Training details of nonlinear DNN with residual connection

The schematic of the nonlinear DNN with residual connections is depicted in Fig. S6a. The system parameters are kept the same as the one shown in Fig.2 of the main text, with an additional beam splitter. Alongside the trainable phase on the SLM, two other parameters are introduced as trainable parameters: the reflectance of the beam splitter and the nonlinear refractive index of the nonlinear medium. When light first passes through the beam splitter, it splits into two beams: one is modulated by the SLM and the other is activated and reflected by the mirror-coated Si wafer. Each subsequent pass through the beam splitter further divides the light. The equivalent electronic neural network is depicted in Fig. S6b, which creates a more intricate residual connection than those in traditional neural networks. The nonlinear DNN with residual connections shows an improved classification accuracy, as depicted in Fig. 4f of the main text, and can be further optimized by replacing the uniform beam splitter by a beam splitter with spatially-varying reflectance.

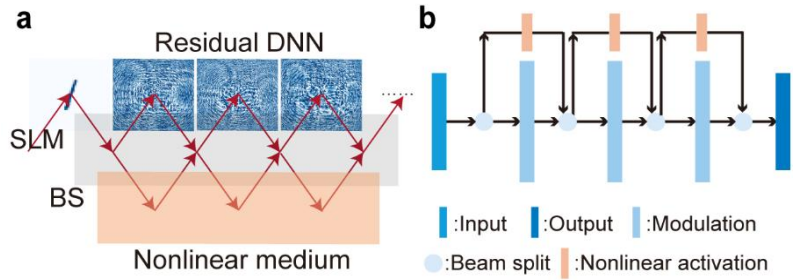

**Figure S6 | a**, Schematic of the scalable nonlinear DNN with residual connection realized by a single uniform beam splitter. **b**, Schematic of the equivalent electronic neural network.

## 9. 25-class QuickDraw dataset

Examples from each category of the 25-class QuickDraw dataset[13] can be visualized in Fig. S6. Unlike standard datasets, the QuickDraw dataset offers greater variability, as each category includes drawings from different individuals, making it a more challenging benchmark.

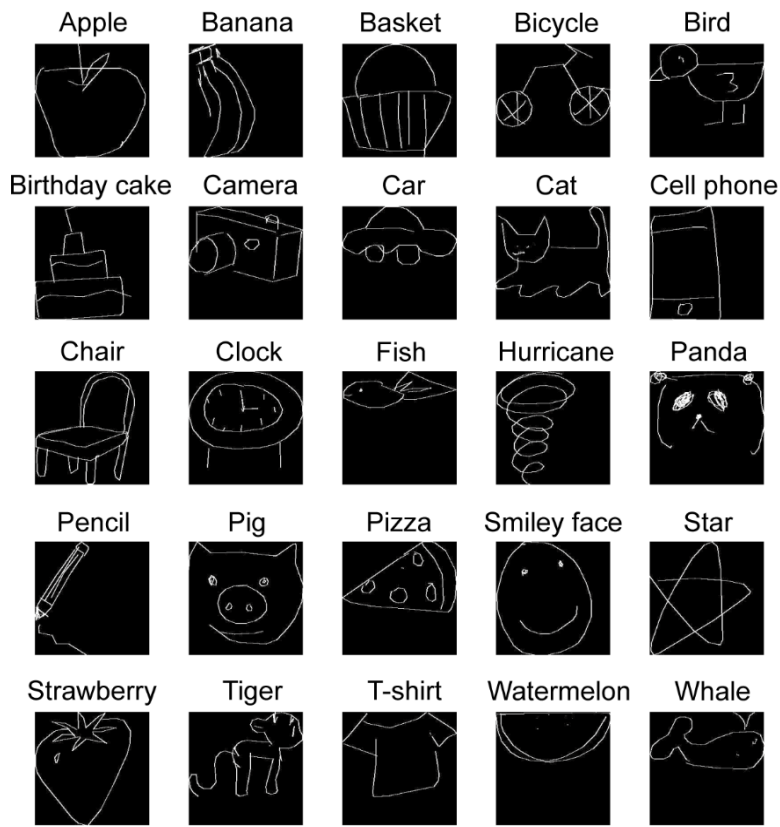

**Figure S6 |** Examples of 25-class QuickDraw dataset.